\documentclass[prd,aps,nofootinbib,floatfix,11pt]{revtex4}

\usepackage{amsmath,graphicx,epsfig,amssymb,dsfont,mathtools}
\usepackage[usenames]{color}
\usepackage{ulem} 
\usepackage{bigstrut}
\usepackage{slashed}
\usepackage{multirow}
\usepackage{subfigure}

\allowdisplaybreaks

\begin{document}
	
\title{A progress in inverse matrix method in QCD sum rules}

\author{
Zhen-Xing Zhao$^{1}$~\footnote{Email: zhaozx19@imu.edu.cn},
Yi-Peng Xing$^{1}$,
Run-Hui Li$^{1}$~\footnote{Email: lirh@imu.edu.cn}
}

\affiliation{
$^{1}$ School of Physical Science and Technology, \\
Inner Mongolia University, Hohhot 010021, China
}
		
\begin{abstract}
In traditional QCD sum rules, the simple hadron spectral density model of ``delta-function-type ground state + theta-function-type continuous spectrum" determines that there is no perfect parameter selection. In recent years, inverse problem methods, especially the inverse matrix method, have shown better handling of QCD sum rules. This work continues to develop the inverse matrix method. Considering that the narrow-width approximation may still be a good approximation, we separate the contribution of the ground state from the spectral density. Then follow the general steps of the inverse matrix method to extract physical quantities such as decay constants that we are sometimes more interested in. 
\end{abstract}

\maketitle

\section{Introduction}

In 1978, Shifman, Vainshtein, and Zakharov published their famous
two articles \cite{Shifman:1978bx,Shifman:1978by}, laying the foundation
for QCD sum rules (QCDSR). Subsequently, this method was widely applied
in many aspects of hadron physics, including but not limited to: determination
of quark mass, calculation of form factors for mesons and baryons,
calculation of hadron spectra and decay constants, calculation of
parameters in some effective theories, and study of properties of
exotic states. Some recent works can be found in Refs. 
\cite{Wang:2024gvf,Wang:2024brl,Wang:2024fwc,Wang:2017uld,Yin:2021cbb,Shi:2023kiy,Hu:2017dzi,Shi:2019hbf,Zhao:2020mod,Zhao:2021lzd,Zhao:2021sje,Xing:2021enr,Sun:2023noo}.
Although QCDSR has achieved great success and has even
been written into standard quantum field theory textbooks such as
that by Peskin and Schroeder, the questioning of this method has never
stopped. For example, as early as about 30 years ago in Ref. \cite{Leinweber:1995fn},
it was mentioned that: ``You can get anything you
want from QCD Sum Rules."

Of course, this is not the case, but it also means that QCDSR is not
flawless. Let us briefly review the specific methods for handling
QCDSR in the literature, as well as their respective shortcomings.
To be specific, one can take the calculation of the $\rho$ meson
decay constant $f_{\rho}$ as an example. 

The traditional method. First, the assumption of quark hadron duality
is introduced to extract $f_{\rho}$, where the continuum threshold
parameter $s_{0}$, based on experience, should be near the first
excited state. Second, determine Borel window based on two criteria:
pole dominance and OPE convergence. Finally, by searching for stability
region in the Borel window, $s_{0}$ is ultimately determined, followed
by the value of $f_{\rho}$. Some comments are in order. (1) One can
easily check that $\partial f_{\rho}^{2}/\partial s_{0}>0$. Therefore,
the overly simplified modeling of hadron spectral density
``delta-function-type ground state + theta-function-type continuous spectrum"
determines from the beginning that there is no perfect parameter selection.
(2) Why demand pole dominance when the contribution of excited states
and continuous spectra has been subtracted through quark hadron duality?
How to quantitatively require OPE convergence? Note that the selection
of Borel window directly affects the final numerical results. (3)
What if the stability region does not exist?

The Monte Carlo method. In Ref. \cite{Leinweber:1995fn}, a Monte-Carlo
based method is introduced. It is essentially an improved version
of the traditional method. Various sources of error can be automatically
considered, but at the cost of a huge increase in computational complexity.
The main issue is still the selection of Borel window -- requiring
the contribution of the pole to be greater than 50\% and that of highest
dimensional operator to be less than 10\%. These numbers are too artificial. 

The moment method. Very recently, Ref. \cite{Carvunis:2024koh} suggested
using the moment method to handle QCDSR without awakening quark hadron
duality. The higher the order $n$ of differentiation,
the smaller the contribution of the excited states and continuous
spectra. However, for a fixed $Q^{2}$, $n$ cannot be arbitrarily
large, otherwise the contribution of the condensate terms will dominate.
If $Q^{2}$ tends to infinity, $n$ can also be arbitrarily large,
while the stability criterion requires that the ratio of the two should
be a constant \cite{Reinders:1984sr}. In this sense, this method is
equivalent to the traditional Borel transformed method. 

The method of variable continuum threshold parameter. By requiring
the Borel parameter dependence matching on both sides of the sum
rule, Ref. \cite{Melikhov:2015qva} introduced the so-called
effective continuum threshold that depends on the Borel parameter.
Consider the threshold parameter as a polynomial of the Borel parameter,
and then fit the hadron mass within the Borel window. 

Some interjected remarks. In some literature, the continuum threshold
parameter $s_{0}$ is determined using the derived sum rule for hadron
mass from the two-point correlation function. In principle, one can
also obtain a formula for hadron mass from the sum rule of the three-point correlation
function -- but unfortunately, no matter how $s_{0}$ is tuned,
it is usually difficult to reproduce the experimental value.

QCD sum rules as inverse problems \cite{Li:2020ejs,Xiong:2022uwj}. Given the
calculation results at the QCD level, solve the hadron spectral
density in dispersion integral. Similar to the continuous threshold
parameter $s_{0}$, in Ref. \cite{Li:2020ejs}, a larger cutoff parameter $\Lambda$ is introduced.
The hadron spectral density model is no longer ``ground state + continuous
spectrum'', but ``ground state + low excited states + continuous
spectrum''. Here, the spectral density of low excited states is obtained
by fitting a large range of $q^{2}$ or the squared Borel mass $M^{2}$.
The fitting results depend significantly on the selection of fitting
intervals. A possibly more rigorous prescription can be found in Ref. \cite{Xiong:2022uwj}.

Inverse matrix method \cite{Li:2021gsx,Li:2022qul,Li:2022jxc,Li:2023dqi,Li:2023yay,Li:2023ncg,Li:2023fim,Li:2024awx,Li:2024xnl}.
This method can be regarded as one of the methods for solving inverse problems.
The unknown spectral density in dispersion integral is expanded using generalized Laguerre polynomials,
and then a matrix equation is established by equating the coefficients of $1/(q^{2})^{m}$ on both sides of the sum rule.
Solve the matrix equation to obtain an approximate solution for the
spectral density. This method nearly does not introduce any assumptions such as quark hadron duality,
and can nicely reproduce the peak of the ground state. However,
the accuracy of the extracted decay constant is poor. In this work,
we point out that with slight modifications to this method, one can
obtain physical quantities that we are sometimes more interested in,
such as decay constants, form factors, and so forth.

This paper is structured as follows. In Sec. II, first, the inverse matrix method is presented,
and then, we propose a modified version of this method to extract the decay constant.
Note that the inverse matrix method here is slightly different from that in Ref. \cite{Li:2021gsx}.
The numerical results are shown in Sec. III. The last section contains summary and outlook. 

\section{Inverse matrix method}

\subsection{$\rho$ meson mass}

Following the standard procedure, one can get the sum rule for the
$\rho$ meson
\begin{equation}
\int_{0}^{\Lambda}ds\frac{\rho^{h}(s)}{s-q^{2}}=c\int_{0}^{\Lambda}ds\frac{1}{s-q^{2}}+\frac{c_{2}}{(q^{2})^{2}}+\frac{c_{3}}{(q^{2})^{3}},\label{eq:SR_old}
\end{equation}
where
\begin{align}
 & c=\frac{1}{4\pi^{2}}\left(1+\frac{\alpha_{s}}{\pi}\right),\nonumber \\
 & c_{2}=\frac{1}{12\pi}\langle\alpha_{s}G^{2}\rangle+2\langle m_{q}\bar{q}q\rangle,\quad c_{3}=\frac{224\pi}{81}\kappa\alpha_{s}\langle\bar{q}q\rangle^{2}.
\end{align}
In Eq. (\ref{eq:SR_old}), $\rho^{h}$ is the hadron spectral density
that we are interested in, and $\Lambda$ is a sufficiently large
cutoff. The hadron spectral density $\rho^{h}$ satisfies the following
boundary conditions:
\begin{equation}
\rho^{h}(s=0)=0,\quad\rho^{h}(s=\Lambda)=c,\label{eq:boundary}
\end{equation}
and therefore, $\rho^{h}$ also depends on $\Lambda$. Since $\rho^{h}$
is dimensionless, it can only be a function of $s/\Lambda$. With
the variable changes $x=q^{2}/\Lambda$ and $y=s/\Lambda$, Eq. (\ref{eq:SR_old})
then reduces to
\begin{equation}
\int_{0}^{1}dy\frac{\rho^{h}(y)}{x-y}=c\int_{0}^{1}dy\frac{1}{x-y}-\frac{c_{2}}{x^{2}\Lambda^{2}}-\frac{c_{3}}{x^{3}\Lambda^{3}}.\label{eq:main_SR_new}
\end{equation}

Expanding $\rho^{h}(y)$ in terms of Legendre polynomials
\begin{equation}
\rho^{h}(y)=\sum_{n=0}^{N+1}a_{n}P_{n}(2y-1),\label{eq:rhoh_Legendre}
\end{equation}
substituting
\begin{equation}
\frac{1}{x-y}=\sum_{m=0}^{N-1}\frac{y^{m}}{x^{m+1}}
\end{equation}
into Eq. (\ref{eq:main_SR_new}), then equating the coefficients of
the $1/x^{m+1}$ term, one can obtain
\begin{equation}
\sum_{n=0}^{N+1}\left(\int_{0}^{1}dy\ y^{m}P_{n}(2y-1)\right)a_{n}=c\frac{1}{m+1}+\cdots,\quad m=0,\cdots,N-1,\label{eq:equations_main}
\end{equation}
where the first ellipsis stands for the contributions from the condensates.
Two additional equations are obtained by the boundary conditions in
Eq. (\ref{eq:boundary})
\begin{equation}
\sum_{n=0}^{N+1}a_{n}P_{n}(-1)=0,\quad\sum_{n=0}^{N+1}a_{n}P_{n}(1)=c.\label{eq:boundary_new}
\end{equation}
Solving these $N+2$ linear equations in Eqs. (\ref{eq:equations_main})
and (\ref{eq:boundary_new}), one can obtain $(a_{0},\cdots,a_{N+1})$,
followed by the spectral density $\rho^{h}$ via Eq. (\ref{eq:rhoh_Legendre}).
Then one can read $m_{\rho}$ from the plot of $\rho^{h}(s)$.

The Legendre polynomials $P_{n}$ satisfy the orthogonality 
\begin{equation}
\int_{-1}^{1}dx\ P_{m}(x)P_{n}(x)=\frac{2}{2n+1}\delta_{mn},\label{eq:orthogonality}
\end{equation}
which leads to $\int_{0}^{1}dy\ y^{m}P_{n}(2y-1)=0$ for $m<n$. Therefore,
the coefficient matrix in Eqs. (\ref{eq:equations_main}) and (\ref{eq:boundary_new})
is almost a lower triangular matrix, except for the last two rows. 

In practice, one must stop at a finite $N$. The optimal choice of
$N$ is set to its maximal value, above which a solution goes out
of control. This is the generic feature of an ill-posed inverse problem.
More details can be found in Ref. \cite{Li:2021gsx}.

\subsection{Decay constant $f_{\rho}$}

One can see from the following Fig. \ref{fig:rho} that, although inverse matrix
method can reproduce the $\rho$ meson mass relatively well,
the reproduced total width is much larger than the experimental value
$\Gamma_{\rho}^{{\rm exp}}\approx0.15\ {\rm GeV}$ \cite{ParticleDataGroup:2022pth}.
Therefore, the narrow-width approximation may still be a good approximation.
Based on this consideration, one can pull out the contribution of
the ground state from $\rho^{h}$ in Eq. (\ref{eq:SR_old})
\begin{equation}
\frac{f_{\rho}^{2}}{m_{\rho}^{2}-q^{2}}+\int_{0}^{\Lambda}ds\frac{\bar{\rho}^{h}(s)}{s-q^{2}}=c\int_{0}^{\Lambda}ds\frac{1}{s-q^{2}}+\frac{c_{2}}{(q^{2})^{2}}+\frac{c_{3}}{(q^{2})^{3}},\label{eq:my_SR_old}
\end{equation}
where the subtracted spectral density $\bar{\rho}^{h}$ is related to
the original $\rho^{h}$ via
\begin{equation}
\rho^{h}(s)=f_{\rho}^{2}\delta(s-m_{\rho}^{2})+\bar{\rho}^{h}(s).
\end{equation}
The following boundary conditions are adopted for $\bar{\rho}^{h}$: 
\begin{equation}
\bar{\rho}^{h}(s=0)=0,\quad\bar{\rho}^{h}(s=\Lambda)=c.\label{eq:boundary2}
\end{equation}
$f_{\rho}^{2}$ is certainly the quantity that interests us. Also
denoting $q^{2}/\Lambda\equiv x$ and $s/\Lambda\equiv y$ as above,
and in addition, $m_{\rho}^{2}/\Lambda\equiv y_{0}$ and $f_{\rho}^{2}/\Lambda\equiv a_{N+2}$,
one can rewrite Eq. (\ref{eq:my_SR_old}) into 
\begin{equation}
\frac{a_{N+2}}{x-y_{0}}+\int_{0}^{1}dy\frac{\bar{\rho}^{h}(y)}{x-y}=c\int_{0}^{1}dy\frac{1}{x-y}-\frac{c_{2}}{x^{2}\Lambda^{2}}-\frac{c_{3}}{x^{3}\Lambda^{3}}.\label{eq:my_SR_new}
\end{equation}

Expanding $\bar{\rho}^{h}(y)$ using Legendre polynomials
\begin{equation}
\bar{\rho}^{h}(y)=\sum_{n=0}^{N+1}a_{n}P_{n}(2y-1),\label{eq:rhohbar_Legendre}
\end{equation}
substituting
\begin{equation}
\frac{1}{x-y_{(0)}}=\sum_{m=0}^{N}\frac{y_{(0)}^{m}}{x^{m+1}}
\end{equation}
into Eq. (\ref{eq:my_SR_new}), then equating the coefficients of
the $1/x^{m+1}$ term, one can obtain 
\begin{equation}
a_{N+2}\,y_{0}^{m}+\sum_{n=0}^{N+1}\left(\int_{0}^{1}dy\ y^{m}P_{n}(2y-1)\right)a_{n}=c\frac{1}{m+1}+\cdots,\quad m=0,\cdots,N.\label{eq:equations_main_2}
\end{equation}
Two boundary conditions in Eq. (\ref{eq:boundary2}) are changed into
\begin{equation}
\sum_{n=0}^{N+1}a_{n}P_{n}(-1)=0,\quad\sum_{n=0}^{N+1}a_{n}P_{n}(1)=c.\label{eq:boundary2_new}
\end{equation}
Solving these $N+3$ equations, one can obtain $(a_{0},\cdots,a_{N+2})$,
especially, 
\begin{equation}
a_{N+2}=f_{\rho}^{2}/\Lambda.\label{eq:f_rho}
\end{equation} 

In the following, a new boundary condition
\begin{equation}
\bar{\rho}^{h}(s=m_{\rho}^2)=0
\end{equation}
will be adopted to replace the first one of Eq. (\ref{eq:boundary2}) to improve the calculation accuracy.

\section{Numerical analysis}

\begin{figure}[!]
\includegraphics[width=1.0\columnwidth]{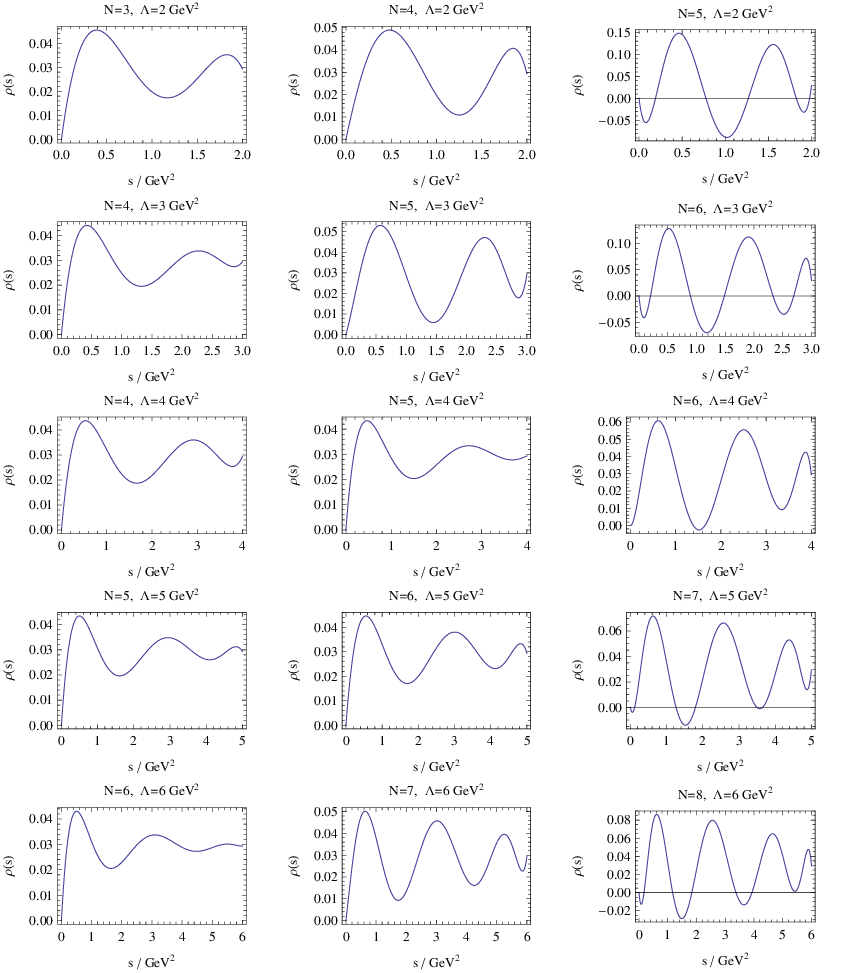}
\caption{The spectral density $\rho^{h}$ in Eq. (\ref{eq:SR_old}) as a function of $s$ with $s\in [0, \Lambda]$, while the $\Lambda$ and $N$ in Eq. (\ref{eq:rhoh_Legendre}) serve as parameters.}
\label{fig:rho} 
\end{figure}
\begin{figure}[!]
\includegraphics[width=0.6\columnwidth]{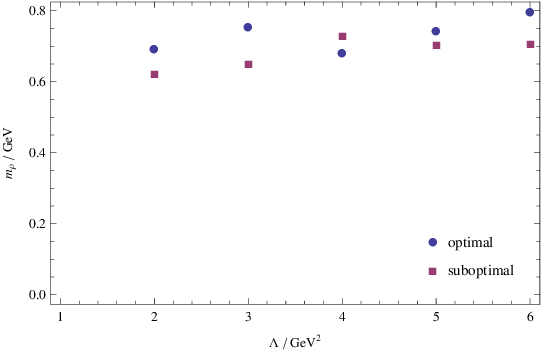}
\caption{$m_{\rho}$ as a function of $\Lambda$. The suboptimal and optimal values of $m_{\rho}$ are read from the first and second columns in Fig. \ref{fig:rho}.}
\label{fig:mrho} 
\end{figure}
\begin{figure}[!]
\includegraphics[width=1.0\columnwidth]{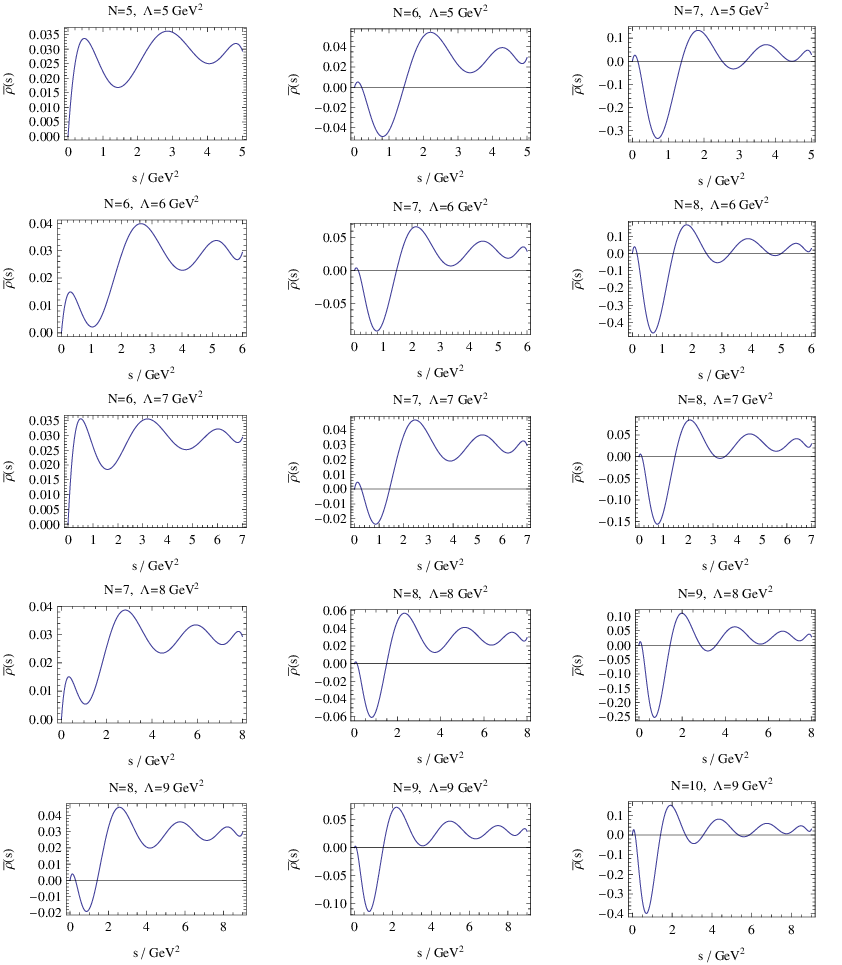}
\caption{The subtracted spectral density $\bar{\rho}^{h}$ in Eq. (\ref{eq:my_SR_old}) as a function of $s$ with $s\in [0, \Lambda]$, while the $\Lambda$ and $N$ in Eq. (\ref{eq:rhohbar_Legendre}) serve as parameters. The boundary condition in Eq. (\ref{eq:boundary1_old}) has been used.} 
\label{fig:rhobar} 
\end{figure}
\begin{figure}[!]
\includegraphics[width=0.6\columnwidth]{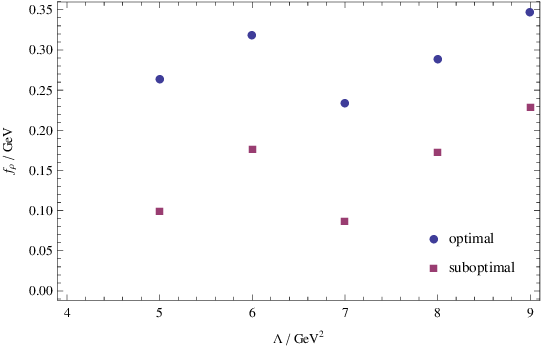}
\caption{$f_{\rho}$ as a function of $\Lambda$. The suboptimal and optimal values of $f_{\rho}$ are read from the first and second columns in Fig. \ref{fig:rhobar}.}
\label{fig:frho} 
\end{figure}
\begin{figure}[!]
\includegraphics[width=1.0\columnwidth]{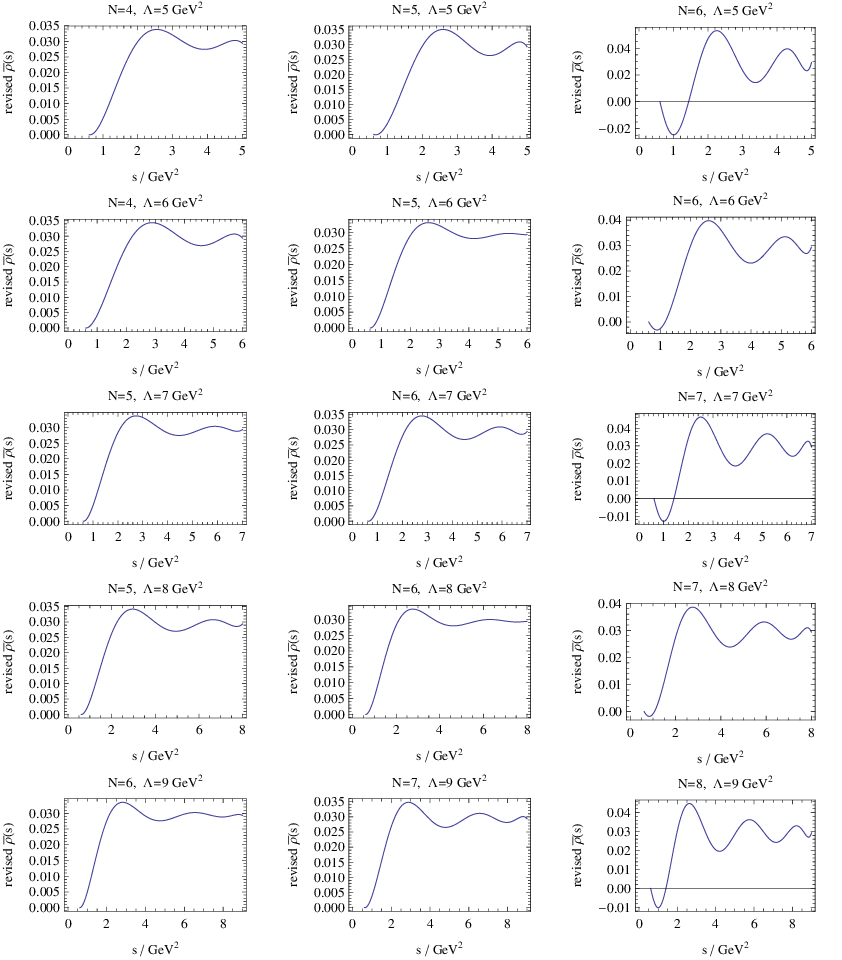}
\caption{The subtracted spectral density $\bar{\rho}^{h}$ in Eq. (\ref{eq:my_SR_old}) as a function of $s$ with $s\in [m_{\rho}^{2}, \Lambda]$, while the $\Lambda$ and $N$ in Eq. (\ref{eq:rhohbar_Legendre}) serve as parameters. The boundary condition in Eq. (\ref{eq:boundary1_new}) has been used.}
\label{fig:rhobar_new} 
\end{figure}
\begin{figure}[!]
\includegraphics[width=0.6\columnwidth]{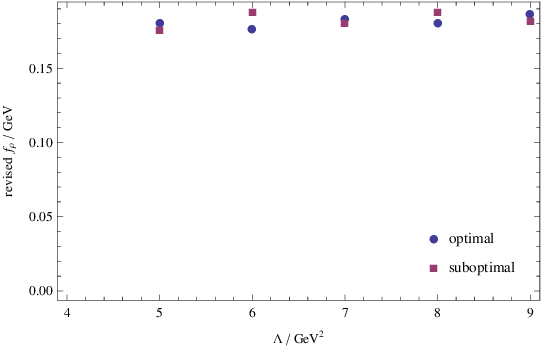}
\caption{$f_{\rho}$ as a function of $\Lambda$. The suboptimal and optimal values of $f_{\rho}$ are read from the first and second columns in Fig. \ref{fig:rhobar_new}.}
\label{fig:frho_new} 
\end{figure}

We take the same OPE parameters and strong coupling as those in Ref. \cite{Li:2021gsx}:
\begin{align}
 & \langle m_{q}\bar{q}q\rangle=0.007\times(-0.246)^{3}\ {\rm GeV}^{4},\quad\langle\alpha_{s}G^{2}\rangle=0.08\ {\rm GeV}^{4},\nonumber \\
 & \alpha_{s}\langle\bar{q}q\rangle^{2}=1.49\times10^{-4}\ {\rm GeV}^{6},\ \alpha_{s}=0.5,\ \kappa=2.
\end{align}

The spectral density $\rho(s)$ can in principle be measured experimentally, and it is positive.
In the following discussion, we will use this fact as the primary criterion.

In Fig. \ref{fig:rho}, we plot the spectral density $\rho^{h}$ in Eq. (\ref{eq:SR_old}) as a function of $s$ with $s\in [0, \Lambda]$, while the $\Lambda$ and $N$ in Eq. (\ref{eq:rhoh_Legendre}) serve as parameters. As can be seen in Fig. \ref{fig:rho} that, the resonances have shown their faces. However, those peaks do not necessarily represent physical states. In the last column of Fig. \ref{fig:rho}, the spectral density has just lost its positivity. Therefore, the middle column corresponds to the optimal solutions of spectral density with respect to $N$, while the first column corresponds to the suboptimal ones. The mass of ground state can be read from the spectral density plots in the first and second columns. The corresponding suboptimal and optimal results are then plotted in Fig. \ref{fig:mrho}. As can be seen in Fig. \ref{fig:mrho} that, the dependence of $m_{\rho}$ on $\Lambda$ is weak around 3-4 ${\rm GeV}^{2}$, where we obtain the optimal estimate of $m_{\rho}$:
\begin{equation}
m_{\rho}=(0.68\text{-}0.76)\,{\rm GeV}.
\end{equation}
Incidentally, when $\Lambda = 3\ {\rm GeV}^{2}$, one might be able to read the approximate location of the first excited state $\rho(1450)$ from Fig. \ref{fig:rho}.

In Fig. \ref{fig:rhobar}, we plot the subtracted spectral density $\bar{\rho}^{h}$ in Eq. (\ref{eq:my_SR_old}) as a function of $s$ with $s\in [0, \Lambda]$, while the $\Lambda$ and $N$ in Eq. (\ref{eq:rhohbar_Legendre}) serve as parameters. As can be seen in Fig. \ref{fig:rhobar} that, a hole is present around $s=m_{\rho}^{2}$. In the last column of Fig. \ref{fig:rhobar}, the subtracted spectral density has just lost its positivity at large $s$. The first and second columns correspond to the suboptimal and optimal solutions of subtracted spectral density with respect to $N$. The corresponding suboptimal and optimal results of the decay constant $f_{\rho}$, which is obtained via Eq. (\ref{eq:f_rho}), are then plotted in Fig. \ref{fig:frho}. As can be seen in Fig. \ref{fig:frho} that, the dependence of $f_{\rho}$ on $\Lambda$ is weak around 6-7 ${\rm GeV}^{2}$, where we obtain the optimal estimate of $f_{\rho}$:
\begin{equation}
f_{\rho}=(0.23\text{-}0.32)\,{\rm GeV}.
\end{equation}
It can be seen that, the prediction of the decay constant is unsatisfactory, as the uncertainty is large and the predicted value is also significantly larger than the experimental value $f_{\rho}^{\rm exp}\approx 0.2 {\rm \,GeV}$.

The reason for the method obtaining a larger decay constant is analyzed as follows: The hole around $m_{\rho}^{2}$ cannot be too deep. It is likely that the first column in Fig. \ref{fig:rhobar} is not yet stable, while the second column has already gone out of control. This also reflects that the accuracy of the method is indeed not high.

In order to improve the calculation accuracy for Eq. (\ref{eq:my_SR_old}) or (\ref{eq:my_SR_new}), we adopt the following new boundary condition
\begin{equation}
\bar{\rho}^{h}(s=m_{\rho}^{2})=0\label{eq:boundary1_new}
\end{equation}
instead of the old one in Eq. (\ref{eq:boundary2})
\begin{equation}
\bar{\rho}^{h}(s=0)=0.\label{eq:boundary1_old}
\end{equation}
After that, in Fig. \ref{fig:rhobar_new}, we replot the subtracted spectral density $\bar{\rho}^{h}$ as a function of $s$ with $s\in [m_{\rho}^{2}, \Lambda]$. Also in the last column of Fig. \ref{fig:rhobar_new}, $\bar{\rho}^{h}$ has just lost its positivity. The first and second columns correspond to the suboptimal and optimal solutions of $\bar{\rho}^{h}$ with respect to $N$. The corresponding suboptimal and optimal results of the decay constant $f_{\rho}$ are then plotted in Fig. \ref{fig:frho_new}. As can be seen in Fig. \ref{fig:frho_new} that, the dependence of $f_{\rho}$ on $\Lambda$ is weak in the full range of 5-9 ${\rm GeV}^{2}$, where we obtain the optimal estimate of $f_{\rho}$:
\begin{equation}
f_{\rho}=(0.18\text{-}0.19)\,{\rm GeV}.
\end{equation}

\section{Conclusions and discussions}

In traditional QCD sum rules, the simple hadron spectral density model of ``delta-function-type ground state + theta-function-type continuous spectrum" determines that there is no perfect parameter selection. In traditional methods, the selection of parameters is not a problem in semi-quantitative aspect, which can be seen from the fact that when setting the threshold parameter to infinity, sum rule of the Borel transformed version or the moment method version yields a decent upper bound for the physical quantity of interest. However, the oversimplified model makes accurate calculation almost impossible. The selection of continuum threshold parameter has become the main source of error in traditional sum rules.

The existing improvements to the traditional method in the literature usually select a so-called Borel window for fitting. However, the fitting results often rely heavily on the selection of the fitting interval. In recent years, inverse problem methods, especially the inverse matrix method, have shown better handling of QCD sum rules. In the inverse matrix method, the unknown hadron spectral density is expanded using a set of orthogonal functions, and then by equating the coefficients of $1/(q^{2})^{m}$ on both sides of the sum rule, the integral equation is converted into a matrix equation. Solve this matrix equation to obtain possible solutions for spectral density.

This work continues to develop the inverse matrix method. Considering that the narrow-width approximation may still be a good approximation, we separate the contribution of the ground state from the spectral density. Then follow the general steps of the inverse matrix method to extract physical quantities such as decay constants or form factors that we are sometimes more interested in. 

The inverse matrix method nearly does not introduce any assumptions such as quark-hadron duality. It only contains few parameters, and more importantly, the extracted physical quantities usually have optimal values with respect to these parameters.

Of course, this method is not perfect and still has some flaws. Most obviously, although this method can reproduce the ground state mass relatively well, the reproduced width is much larger than the experimental value. That is to say, the obtained resonance peak is too ``fat". In addition, when extracting the decay constant in this article, the ground state mass was selected as the input. Is it possible to obtain both the mass and decay constant of the ground state at the same time? We will consider these issues in our future work.

\section*{Acknowledgements}

The authors are grateful to Prof. Hsiang-nan Li for valuable discussions.
This work is supported in part by scientific research start-up fund for
Junma program of Inner Mongolia University, scientific research start-up
fund for talent introduction in Inner Mongolia Autonomous Region,
and National Natural Science Foundation of China under Grants No.
12065020, 12075126.

\end{document}